\documentstyle[preprint,aps]{revtex}
\tighten
\begin{document}
\draft

\title{\bf TWO-PARTICLE RESONANT STATES IN A MANY-BODY MEAN FIELD}

\author  {\rm R. Id Betan $^{a,b)}$,
              R. J. Liotta $^{a)}$,
              N. Sandulescu $^{a,c)}$,
              T. Vertse $^{a,d}$}
\bigskip

\address {\rm
  $^{a)}$~  Royal Institute of Technology, SCFAB,
 SE-10691, Stockholm, Sweden \\
  $^{b)}$~  Departamento de Fisica, FCEIA, UNR,
 Avenida Pellegrini 250,\\ 2000 Rosario, Argentina\\
  $^{c)}$~ Institute of Physics and Nuclear Engineering, \\
P.O.Box MG-6, Bucharest-Magurele, Romania\\
  $^{d)}$~ Institute of Nuclear Research of the 
Hungarian  Academy of Sciences,\\
H-4001 Debrecen, Pf. 51, Hungary}

\maketitle
\begin{abstract}
A formalism to evaluate the resonant states produced 
by two particles moving outside
a closed shell core is presented.
The two particle states are calculated by using a single
particle representation consisting of bound states, Gamow resonances
and scattering states in the complex energy plane (Berggren 
representation).
Two representative cases are analysed corresponding to whether the Fermi 
level is below or above the continuum threshold. 
It is found that long lived two-body states (including bound states)
are mostly determined by either bound single-particle states or by 
narrow Gamow resonances. However, they 
can be significantly affected by the continuum part of the spectrum.
\end{abstract}


The prospect of reaching and measuring very unstable nuclei, as
is materializing now, opens the possibility of studying 
spectroscopic processes occuring in the continuum
part of nuclear spectra. Much work has already been done in 
this subject, particularly regarding halo nuclei \cite{mar}. Still,
the role played by single-particle resonances and of the
continuum itself upon particles moving in the continuum
of a heavy nucleus is not fully understood.
For instance, one may wonder whether 
two particles outside a core where the Fermi level is immersed
in the continuum may produce a quasibound state and, in this
case, whether that state is built upon narrow 
single-particle resonances or by an interplay
between the two-particle interaction and the continuum, 
or by a combination of these mechanisms, as it happens in
typical halo nuclei.  
To answer such questions is a difficult undertaking, 
particularly because the resonances on the real energy
axis do not correspond to a definite state. 
A way of approaching this problem is by 
solving the Schr\"odinger equation with
outgoing boundary conditions.
One thus obtains the resonances as poles of the S-matrix
in the complex energy plane. These poles (Gamow resonances)
can be considered discrete states on the same footing as 
bound states (see Ref. \cite{lio} and
references therein). 
However, in this case one finds that
physical quantities, like energies and probabilities, become
complex. One may attempt to give meaning to these complex 
quantities. Thus, it is usually assumed that the
imaginary  part of the energy of a decaying resonance
is (except a minus sign) half the
width.  Other examples are
the interpretation of complex cross sections done  by Berggren
\cite{tb} or the widely used
radioactive decay width evaluated by Thomas as the
residues of the S-matrix \cite{th}.
All these interpretations are valid only if the
resonances are isolated and, therefore, narrow.
In this case the residues
of the S-matrix becomes real.  
One may thus apply this theory and evaluate all resonances, 
giving physical meaning to the narrow ones only. 
To achieve this goal a representation consisting of 
bound states, Gamow resonances and the proper continuum was 
proposed some years ago \cite{lio} (Berggren representation). 
One chooses the proper
continuum as a given contour in the complex energy plane 
and forms the basis set of states (the representation) as 
the bound states plus the Gamow resonances included in that 
contour plus the scattering states on the contour \cite{lio}. 

Using the Berggren representation one 
can evaluate any one-particle
quantity in the complex energy-plane, e. g. 
the eigenstates of a deformed potential in 
terms of the Berggren states provided by
a spherical basis. One may 
thus think that the Berggren representation 
can also be used straightaway  to evaluate many-particle 
quantities, as one does with the shell-model using bound 
representations. Unfortunately this is not the case. The
root of the problem is that the set of energies of the two-particle basis 
states may cover the whole complex energy plane of interest.
To show this we will analyse the relatively simple 
case of two particles outside a core in terms
of the Green function. For clarity of presentation we will
give here the main points of the derivations leading to the Berggren 
representation.

The single-particle Green function in the
complex energy plane can be written as \cite{be93}
\begin{equation}\label{eq:spgf}
g(r,r^{'};E)=\sum_{n=1}^{N_d}
{w_n(r)w_n(r^{'})\over{E-\epsilon_n}} +
\int_{L^+} d\epsilon{u(r,\epsilon)u(r^{'},\epsilon)
\over{E-\epsilon}}
\end{equation}
where $L^+$ is an integration path on the complex energy
plane and $u(r,\epsilon)$ is the corresponding scattering 
function. The index $n$ labels the bound states and Gamow resonances 
which lie between the real energy axis and the integration path and
$w_n(r)$ are the corresponding wave functions, i. e. 
the outgoing solutions of the Schr\"odinger equation.
The total number of bound plus Gamow states, i. e. of discrete
states, is $N_d$. Notice that the Berggren 
metric in Eq. (\ref{eq:spgf}) corresponds to the product of a
function times itself, and not times its complex conjugate. This
induces the complex probabilities mentioned above. But on the
real energy axis (where the scattering functions can be chosen
to be real) 
the Berggren and Hilbert internal products coincide.
Therefore, since this formalism is based upon a 
Cauchy transformation to the complex energy plane,
it should provide at the same time
the physical complex quantities mentioned above plus the real  
ones as evaluated by using bound representations, e. g.
within the shell model. This is
a strong constraint that will help us to probe the 
formalism and check the computer codes.

The numerical evaluation of the integral requires the 
discretization of the complex energy $\epsilon$
along the integration contour. We will use
the Gauss integration method with a total number of 
integration points $N_g$ and weights $h_p$.
This defines the Berggren representation, i. e. the set of 
orthonormal (in the Berggren metric) 
basis vectors $\{\vert \varphi_j\rangle\}$ given by
the set of bound  and Gamow states
$\{\varphi_p(r)\}=\{w_p(r)\}$ and the discretized 
scattering states
$\{\varphi_p(r)\}=\{\sqrt{h_p}u(r,\epsilon_p)\}$.
The single particle Green function can then be written as
\begin{equation}\label{eq:bspgf}
g(r,r^{'};E)=\sum_{p=1}^N
{\varphi_p(r)\varphi_p(r^{'})
\over{E-\epsilon_p}}
\end{equation}
where $N=N_d+N_g$.

The two-particle Green function is \cite{fw}, 
\begin{equation}\label{eq:tpgf}
G(r,r^{'};E)=G_0(r,r^{'};E) + 
\int dr_1dr_2 G_0(r,r_1;E)V(r_1r_2)G(r_2,r^{'};E) 
\end{equation}
where $G_0$ is the bare (zeroth-order) two-particle Green function which within
the Berggren representation reads,
\begin{equation}\label{eq:btpgf}
G_0(r,r^{'};E) = \sum_{i\leq j=1}^N{\varphi_i(r)\varphi_j(r)
\varphi_i(r^{'})\varphi_j(r^{'})\over E-(\epsilon_i+\epsilon_j)}
\end{equation}
As seen from Eq. (\ref{eq:tpgf}), in order 
to obtain the poles and residua (i. e. 
energies and wave functions) of the correlated two-particle Green 
function one has to assume that they do not coincide
with those in the zeroth-order Green function \cite{fw}. 
But by choosing an arbitrary path for the one-particle scattering
states in the complex energy plane, one may obtain a continuum
plane of zeroth-order poles corresponding to the sum 
$\epsilon_i+\epsilon_j$. Therefore, in this plane one would not be 
able to evaluate any two-particle pole.

As an illuminating example we show in Fig. 
\ref{cont}(a) a rectangular one-particle contour without any
discrete state. From Eq. (\ref{eq:btpgf}) 
one sees that in the complex two-particle
energy plane the zeroth-order energies are given by
the geometrical
sum of a point $i$ on the one-particle contour (corresponding to the complex
energy $\epsilon_i$) plus 
another point $j$ on that contour, such that $i\leq j$. 
These energies are located in
the two-particle complex energy plane as
shown in Fig. \ref{cont}(b).
For example, the dashed region located between $a$ and $2a$
on the real energy axis is produced by one of the particles
lying in the first segment of the one-particle contour, i. e.
with energy $\epsilon_i=(x,0)$ such that $0<x<a$, and the second particle 
in the segment with energy $\epsilon_j=(a,y)$ such that 
$-c<y<0$. Therefore the zeroth-order two-particle energy is
$\epsilon_i+\epsilon_j=(x+a,y)$, which covers the region mentioned 
above. The rest of the dashed plane in Fig. \ref{cont}(b) is produced
in the same fashion, with the energies of the particles on 
different segments of the one-particle contour.

We thus see that if $2a<b$ then there is a region in the two-particle
complex energy plane which is free from any uncorrelated solution.
Therefore, in this region 
one can search for resonant states of the interacting two-particle system. 
By choosing the real energies $a$ and $b$ conveniently, i. e. such that
$2a<b$ as in Fig. \ref{cont}(b), one can study 
two-particle resonances lying in any reasonable energy region. We will  
call this the "allowed" energy region.

The allowed region in the two-particle complex
energy plane plays a role similar to the contour in 
the one-particle energy plane. That is, only those
one-particle resonances with complex energies
that lie between the real energy axis
and the contour can be evaluated by using the Berggren 
representation \cite{lio}. In the same fashion, only those
two-particle resonant states with complex energies
contained in the allowed 
region can be evaluated within that representation.

The allowed region can be determined by fulfilling some
physically meaningful requirements. Thus, one expects that the
single-particle resonances and the continuum scattering states
closest to threshold would play an important role in the
building up of correlated low lying two-particle states. 
Therefore they should be included in the basis, which 
implies that the one-particle contour of Fig. \ref{cont}(a)
should correspond to small values of $a$ and large values of $b$,
as indeed is the case in Fig. \ref{cont}(b).

Using the Berggen representation one can obtain the two-particle
TDA equations in a standard fashion. We will use in our derivations
separable forces so that those equations convert into the usual
dispersion relation \cite{al,epml}, i. e.
\begin{equation}\label{eq:disp}
-1/G_\lambda=\sum_{i\leq j}{C^2(ij,\lambda)
\over{E-\epsilon_i-\epsilon_j}}
\end{equation}
where $E$ is the complex energy of
the two-particle states carrying angular momenta $\lambda$,
$C(ij,\lambda)$ is the $\lambda$-multipole component
of the interaction while $i$ and $j$ label the single-particle states 
with complex energies $\epsilon$ corresponding
to our Berggren representation. For a detailed expression of
the coefficient $C(ij,\lambda)$ see, e. g., Eq. (32) of Ref. \cite{epml}.
It contains the matrix element of the radial field 
$f_\lambda(r)$ in the separable interaction, i. e.
$\int r^2 dr \varphi_i(r) f_\lambda(r) \varphi_j(r)$
Notice again the Berggren metric in the radial internal
products and that it is the square 
(and not the absolute value square) of the
matrix elements that appear in the dispersion relation. 

To generate the single-particle states
we will use a Woods-Saxon potential (WS). 
The field $f_{\lambda}$ is the derivative of the WS.
It may be argued that this interaction is too simple
to describe the motion of the particles in the continuum and
that a more realistic force should be used, as it was done, e. g.,
in Ref. \cite{be}. However, our purpose here is not to explain in
detail processes happening in the continuum, but rather to understand 
the role played by the various ingredients entering into the 
two-particle quasibound state that may be built as a result
of the interplay among those ingredients. We assume that 
this process occurs near the nuclear surface and,
therefore, our separable force should be suitable for the
analysis that we intend to carry out.

We will apply the formalism presented here
to analyse neutron excitations in a nucleus that would lie on or
even beyond the drip line. That is, the Fermi level may be close to
or even immersed in the continuum. We will analyse these two 
possibilities separately. 

The calculation of the bound states and the Gamow
resonances will be performed by using an
updated version of the 
compute code GAMOW \cite{tv1} while the scattering waves on
the complex contour of Fig. \ref{cont}(a) will be evaluated by
using the computer code ZSCAT \cite{tv2}.

The WS to be used correspond
to the double closed shell nucleus $^{78}$Ni.
The 
parameters for the volume part of the interaction are
$V_0=40 MeV, r_0=1.27 fm, a=0.67 fm$. The spin-orbit interaction has
the same values of $r_0$ and $a$, but the depth of the potential is
$V_{so}=21.43 MeV$. 
With these parameters one obtains the single-particle
states shown in Table \ref{sps} under the label $WS1$. 
They are quite similar to the ones given by a Skyrme-HF
calculations \cite{nic}.
The shell $N=50$ is well defined, since there is a gap of
about 3.6 MeV between the lowest particle state, which here
is $1d_{5/2}$, and the highest hole state, i. e. $0g_{9/2}$.

We will also evaluate a case where no bound single-particle
states are present. For this, we reduced the value of the 
depth of the WS to $V_0=37 MeV$. The corresponding
single-particle states are shown under the column WS2
in Table \ref{sps}. One sees that with this
rather shallow potential the gap corresponding
to $N=50$ is still present, given credibility
to our TDA calculation. It is also to be noticed
that the particle state $2s_{1/2}$ has disappeared,
as expected for neutron excitations.

We will here present two-particle states with angular momentum
$\lambda=0$, for which
the separable force is known to reproduce well experimental
data when available.
We will first analyse the case where there are 
bound single-particle states, i. e. the case
$SW1$ in Table \ref{sps}. 
To determine the strength of the separable force we
will follow the standard procedure of adjusting 
$G_\lambda$ by fitting the energy of a two-particle
state, which usually is experimentally known. In our
case we will assume that such state, which
would be the ground state of $^{80}$Ni, exists below
twice the energy of the lowest single particle state,
i. e. below 2$\epsilon_{1d_{5/2}}$. This energy
gap, i. e. the correlation energy, is more than 
1 MeV in well established normal nuclei, like
$^{208}$Pb (where it is 1.244 MeV) and $^{56}$Ni 
(1.936 MeV). However in our case the bound states are
so few and so slightly bound that such high energy gaps
do not seem to be reasonable. Since there is not any experimental
data which could guide us, and since our intention is just 
to see how the strength of the force affects the results,
we will vary the gap from a value of only 300 keV 
to the rather large value of 1.7 MeV  
to examine the differences. 
 
The values of $G_\lambda$ thus obtained depends upon the number
of states included in the basis, as can be seen from 
Eq. (\ref{eq:disp}). In the calculations to be presented here
we used a rectangular contour with the vertices as in Fig.
\ref{cont}(a) with a=0.5 MeV, b=9 MeV, c=-4 MeV and d=20 MeV. 
We thus include in the Berggren basis
all the bound and Gamow states shown in Table
\ref{sps} (except, of course, the hole state $0g_{9/2}$).
The allowed region, therefore, comprises the two-particle
energy plane with complex energies $(E_r,E_i)$ such that
$1 MeV <E_r< 9 MeV$ and $-4 MeV <E_i< 0 MeV$.
 
As in Ref. \cite{lio} we use a Gaussian method of integration
over the contour. The corresponding Gaussian
points provide the scattering waves constituting the 
basis elements on the continuum. We 
have found that in order to obtain 
convergence within six digits in the evaluated quantities,
one has to include
10 Gaussian points for each MeV on the lines of the contour,
except for the last segment 
(the one going from (b,0) to (d,0)) where 5 points for
each MeV is enough.
We arrive to this conclusion
by always choosing the contour such that the resonances lie at least
300 keV from the borders of the contour. The number of scattering
states thus included in the basis is $N_g$=225. In Table \ref{con}
we show the convergence of the results as a function of $N_g$
as well as the influence of the continuum upon the calculated
states. We will come back to this point below.

One can check the reliability of the results by performing a calculation over
the real energy axis only \cite{be}. The real (bound) energies thus 
obtained, which we call "exact", should coincide with those evaluated by using
any contour. Moreover, the value of $G_\lambda$ should, in all cases,
be a real quantity. All these requirements are indeed fulfilled
in our calculations. 

In Fig. \ref{lines} we present all the calculated 
energies which we found inside the allowed region of the
complex two-particle plane. The strength of
the separable force was evaluated assuming 
that the energy gap is 1.4 MeV, i. e. the ground state
energy is -3 MeV.

The first feature that strikes the eye in this figure
is the straight line pattern that follow most of the energy
points. These lines
correspond to basis states where one of the 
particles moves in a bound or Gamow state
and the other in a fragment of the one-particle contour.
For instance, the straight line at a real energy of 
3.796 MeV  corresponds to a particle in the Gamow state
$0h_{11/2}$, with an energy (3.296,-0.013) MeV (as seen in Table
\ref{sps}), while the other is in the $h_{11/2}$ scattering states 
lying on the border at a=0.5 MeV in the contour of Fig. \ref{cont}(a).
The sum of both single-particle energies yields a real part 
of 3.796 MeV, which shows that these states are in fact poles of
the zeroth-order Green function.
Similar structures are found for all the straigth lines in this
figure, with the single-particle quantum numbers 
as indicated in the end of the lines. Thus, the 
horizontal segment at -0.479 MeV corresponding to the 
configuration $d_{3/2}^2$ is produced by a particle in
the Gamow state $1d_{3/2}$ while the other is on the
scattering states belonging to the segment of the contour
on the  real energy axis between 0 and a=0.5 MeV.
This horizontal segment does not appear in the lines labelled
$d_{5/2}^2$ and $s_{s/2}^2$ in Fig. \ref{lines} 
because this lines are generated by the bound single-particle
states coupled to the scattering states on the border of the countour 
lying between (b,0) and (b,-c) in Fig. \ref{cont}(a).

We found that all the lines in Fig. \ref{lines} correspond to 
zeroth-order poles. Therefore the states on the lines are solutions
of both the correlated and the uncorrelated two-particle Hamiltonian. 
They do not describe the physical resonances that we are 
searching and can be considered spurious states.
This peculiar feature of the continuum is also found in the 
one-particle case, where the states lying on the contour
are solutions of both the correlated and the uncorrelated
one-particle Hamiltonian, as it was shown in Ref. \cite{lio}.

Besides these peculiar lines the only two-particle states
inside the allowed region are those indicated by
open circles. 
These states include the two bound states at -3 Mev and -0.653 
MeV as well as a number of
resonances. They are mainly generated by configurations where
both particles occupies bound states and/or Gamow resonances.  
To show the effect of the interaction upon these states, which
are the physical ones, we present in Fig. \ref{bound} the 
corresponding energies as a function of the energy
gap that defines $G_\lambda$.
The surprising feature in this figure is that all the states
become narrower as the interaction increases, except the
state that in zeroth-order is the narrowest one, i. e. 
$h_{11/2}^2$.

The wave function amplitude corresponding
to a given configuration is in our case proportional
to the degeneracy of the configuration. Therefore
the configuration $h_{11/2}^2$ should be decisive in
the building up of narrow two-particle resonances, both
because it is the one with largest degeneracy and also because
in zeroth-order it is the narrowest state. Another feature
of our separable force is that that wave function amplitude
becomes more important as the  
correlated energy approaches the configuration
(zeroth-order) energy. 
This indeed happens in our case, as can be 
seen in Table \ref{wfb}, where the value of the wave function amplitude
$X((0h_{11/2})^2;E)$ is given for the resonances in Fig. \ref{bound}. 
One sees that as the states become narrower the shell $0h_{11/2}$
becomes more important in the corresponding wave functions. And
the other way around: the states labelled
$h_{11/2}^2$ becomes wider
as the interaction increases while the shell $0h_{11/2}$ becomes
less important. 

An important conclusion that can be drawn from Fig. \ref{bound}
is that due to the two-particle interaction
wide resonances can give rise to narrow ones. This is shown
by the states $g_{7/2}^2$ and $d_{3/2}^2$, 
although here one sees that the 
width of the resonance diminishes to reach a minimum
value at $\Delta$ = 1.4 MeV and after that it starts to increase
again. Even the states $g_{7/2}^2$ and $f_{7/2}^2$ reach a
point where increasing $\Delta$ does not affect the energies
much. A result of this is that the corresponding wave function
components $X((0h_{11/2})^2;E)$ also remain unchanged, as
seen in Table \ref{wfb}. The bound states behave in a standard
shell model fashion, as expected for bound states. In particular
the states labelled $s_{1/2}^2$, where the low degenaracy shell 
$2s_{1/2}$ is dominant, are weakly affected by the interaction.

Although the physical states presented above are mainly 
determined by discrete states,
the continuum part of the spectrum plays also an important role.
In particular, one can see in Table \ref{con} that the 
energies evaluated by excluding the scattering states do not 
fit well the correct results.
It is also important to mention that in our calculations we require
the Hamiltonian to be Hermitian, which implies that the strength
$G_\lambda$ has to be a real quantity.\footnote{Due to the 
Berggren metric the matrix representation of the Hamiltonian
is not Hermitian in the complex energy sector.} But by fitting the energy
of $^{80}$Ni(gs) to the value -3 MeV (i. e. $\Delta$ = 1.4 MeV)
and  excluding the continuum,
$G_\lambda$ becomes complex. By taking the corresponding
real part only, as was done in Table \ref{con}, that ground state energy 
acquires the unphysical value (-2.856,0.359) MeV.
Even the energy of the (bound) first excited state $E_1$ is unphysical
since its energy is not real and the 
first resonance, i. e. $E_2$, is unphysical because the 
imaginary part of the energy is positive.
But, as seen in this table, a rather small number
of scattering states is enough to obtain reasonable values for
the energies. Thus, at $N_g$ = 35 one already reaches a presicion
of the order of a few keV.

We have also performed similar calculations by using the single-particle
states labelled WS2 in Table \ref{sps}, where the Fermi level is
immersed in the continuum. Since the resonances are wider than before
we used here a different one-particle contour, namely 
a=0.1 MeV, b=13 MeV, c=-6 MeV and
d= 26 MeV. 

The straight lines discussed above appear
also in this case with the same characteristics as before. 
The remaining physical 
two-particle energies are shown in Fig. \ref{unbound}.
The strengths $G_\lambda$ in the figure are 10 \% larger than 
the corresponding ones in the previous case.

The general features of the states in the continuum
in this figure are similar to those in the 
previous one, as expected since these states are determined mainly by
the Gamow resonances and the continuum background. In fact there is 
not any essential difference between the two calculations since within
this formalism all states (including the continuum states)
are treated on the same footing, indepedently 
of the location of the Fermi level. 
But due to the different single-particle states that enter in the 
calculation the bound states show an striking difference with the 
previous case. Thus, since there is not any bound single-particle 
state and the state $2s_{1/2}$ is not present in this case,
there is only one two-particle bound state which materializes 
only when the interaction is large enough. This occurs in the 
figure at $G=$ 0.023 MeV, where that bound state appears at an energy
of -0.104 MeV. The main components of the  
corresponding wave function are $(-0.95,0.03) (1d_{5/2})^2 +
(0.21,0.00) (0h_{11/2})^2 + (-0.17,0.07) (1d_{3/2})^2 +
(-0.13,0.01) (0g_{7/2})^2 +
(0.11,-0.05) (1f_{7/2})^2$. The interesting point is that
this wave function does not change much as the interaction is increased,
which shows the role played by the Gamow states in building
up the bound states. The importance
of these resonances is related to their widths. The 
wider the Gamow resonance the smaller is their influence. 
However, one cannot conclude from this that only Gamow resonances
would be enough to describe the two-particle states of interest, 
since the
inclusion of the complex contour is important to obtain even
the narrow resonances and the bound states. In particular,
without the contour the imaginary part of the energy 
corresponding to bound states becomes large, as it happened 
in the previous example.

In conclusion we have presented in this paper a method
to perform shell model calculations in the continuum. 
We have shown that wide resonances and even the continuum 
background can be important to describe narrow two-particle
resonances. We thus think that we have solved the old 
problem of describing microscopically resonances induced
by a two-body interaction in the background of a many-body
mean field.

\acknowledgments

This work has been supported by FOMEC (Argentina), by
the Hungarian OTKA fund Nos. T26244 and T29003, by
the Swedish Foundation for International Cooperation
in Research and Higher Education (STINT) and by 
the Swedish Institute.

\begin{figure}
\caption{(a) Rectangular one-particle contour in the complex
energy plane. The points in this contour define the scattering
functions that form the representation to be used in the 
two-particle basis. (b) Continue set of states in the two-particle 
energy plane (dashed region). The white area corresponds to the
allowed region. The values of $a$, 
$b$, $c$ and $d$ are as in the one-particle contour of the previous 
figure.}
\label{cont}
\end{figure}
\begin{figure}
\caption{Poles in the two-particle energy plane corresponding to 
states $\lambda$=0 in $^{80}$Ni. The labels
of the straight lines correspond to configurations in which one
of the two particles is in a bound state or a Gamow resonance 
and the other is on a scattering state. Energies are in MeV.} 
\label{lines}
\end{figure}
\begin{figure}
\caption{Physically meaningful resonances in the two-particle energy plane 
plotted as a function of the energy gap $\Delta$ (in MeV). 
The labels in each group of states
indicate the zeroth-order configuration (i. e. at $\Delta$ = 0 MeV)
corresponding to the group. Energies are in MeV.} 
\label{bound}
\end{figure}
\begin{figure}
\caption{Poles in the two-particle energy plane corresponding to 
the case where the Fermi level lies in the continuum. The value
of the strength $G_\lambda$ (in MeV) was chosen as explained in the tex. 
The energies $(E_r,E_i)$ defining the allowed region are constraint 
to the values 
$0.2 MeV <E_r< 13 MeV$ and $-6 MeV <E_i< 0 MeV$.
The labels in each group of levels indicate the zeroth-order
configuration (i. e. at $G$ = 0) corresponding to the group.
Energies are in MeV.}
\label{unbound}
\end{figure}
\begin{table}
\caption{Neutron single-particle states evaluated with the Woods-Saxon 
potential given in the text. The complex energies are in MeV.
The column labelled $WS1$ corresponds
to $V_0 = 40$ MeV while $WS2$ to $V_0=37$ MeV.
The states $0g_{9/2}$ are given to show the magnitude of the
gap corresponding to the magic number $N=50$.
\label{sps}}
\begin{tabular}{cccccccc}
state&WS1&WS2\\
$0g_{9/2}$&(-4.398,0)&$(-2.587,0)$\\
$1d_{5/2}$&(-0.800,0)&(0.294,-0.018)\\
$2s_{1/2}$&(-0.284,0)&$-----$\\
$1d_{3/2}$&(1.325,-0.479)&(1.905,-1.241)\\
$0h_{11/2}$&(3.296,-0.013)&(4.681,-0.069)\\
$1f_{7/2}$&(3.937,-1.796)&(4.455,-2.851)\\
$0g_{7/2}$&(4.200,-0.167)&(5.799,-0.506)\\
\end{tabular}
\end{table}
\begin{table}
\caption{Energies (in MeV)
of the $\lambda=0$ first excited bound state
and of the lowest two-particle resonances in $^{80}$Ni. 
The discrete single-particle states are those labelled WS1 in
Table \protect\ref{sps}. The
strength $G_\lambda$ was chosen such that the energy gap is 1.4 MeV.
The energies are given as a function of the number of scattering
states included in the single-particle representation, i. e.
the number $N_g$ of Gaussian points. For $N_g$=0 the representation
consists of bound states and Gamow resonances only.
The case $N_g$=225 corresponds to the one used throughout 
the calculations presented here.
\label{con}}
\begin{tabular}{cccccccc}
$N_g$&$E_1$&$E_2$&$E_3$&$E_4$\\
0&(-0.642,0.012)&(2.158,0.719)&(3.268,-0.883)&(7.931,-0.198)\\
35&(-0.65417,0)&(1.96874,-0.39235)&(3.92420,-1.05208)&(7.95693,-0.25236)\\
70&(-0.65274,0)&(1.96988,-0.39321)&
(3.92429,-1.05159)&(7.95691,-0.25251)\\
110&(-0.65274,0)&(1.97261,-0.39838)&
(3.92416,-1.05168)&(7.95687,-0.25250)\\
225&(-0.65308,0)&(1.97241,-0.39935)&
(3.92390,-1.05189)&(7.95685,-0.25249)\\
550&(-0.65308,0)&(1.97241,-0.39935)&
(3.92390,-1.05189)&(7.95685,-0.25249)\\
\end{tabular}
\end{table}
\begin{table}
\caption{Wave function amplitude $X((0h_{11/2})^2;E)$
corresponding to the resonances as labelled in Fig. 
\protect\ref{bound}. The energies $E$ can be read from that figure.
The values of the gap $\Delta$ are given in MeV.
\label{wfb}}
\begin{tabular}{cccccccc}
State&$\Delta=0$&$\Delta=0.6$&$\Delta=1.2$&$\Delta=1.4$&$\Delta=1.7$\\
$d_{3/2}^2$&(0,0)&(0.20,-0.15)&(0.44,-0.08)&(0.48,-0.04)&(0.49,-0.00)\\
$h_{11/2}^2$&(1,0)&(0.92,0.02)&(0.78,0.05)&(0.75,0.03)&(0.71,0.00)\\
$g_{7/2}^2$&(0,0)&(-0.35,-0.02)&(-0.41,-0.02)&(-0.42,-0.02)&(-0.43,-0.02)\\
$f_{7/2}^2$&(0,0)&(-0.16,0.00)&(-0.20,0.02)&(-0.20,0.03)&(-0.21,0.03)\\
\end{tabular}
\end{table}

\begin{thebibliography}{90}

\bibitem{mar}
F. M. Marqu\'es et. al., Phys. Rev. C {\bf 64}, 61301(R) (2001),
and references therein.

\bibitem{lio}
R. J. Liotta, E. Maglione, N. Sandulescu and T. Vertse,
Phys. Lett. {\bf 367B}, 1 (1996).

\bibitem{tb}
T. Berggren, Phys. Lett. {\bf B73}, 389 (1978).

\bibitem{th}
R. G. Thomas, Prog. Theor. Phys. {\bf 12}, 253 (1954);
R. G. Lovas et. al., Phys. Report {\bf 294}, 265 (1998).

\bibitem{be93}
T. Berggren and P. Lind, Phys. Rev. {\bf C47}, 768 (1993).

\bibitem{fw}
A. L. Fetter and J. D. Walecka, Quantum theory of many-particle
systems, McGraw-Hill, New York, 1971.

\bibitem{al}
A. M. Lane, Nuclear theory, Benjaming Reading, Mass, 1964.

\bibitem{epml}
A. Evans et. al., Nucl. Phys. {\bf A93}, 261 (1967).

\bibitem{be}
G. F. Bertsch and H. Esbensen, Ann. of Phys. {\bf 209}, 327 (1991).

\bibitem{tv1}
T. Vertse, K. F. P\'al and Z. Balogh, Comput.
Phys. Commun. {\bf 27}, 309 (1982).

\bibitem{tv2}

L. Gr. Ixaru, M. Rizea and T. Vertse, 
Comput. Phys. Commun. {\bf 85}, 217 (1995).

\bibitem{nic}
N. Sandulescu, N. Van Giai and R. J. Liotta,
Phys. Rev. C {\bf 61}, 61301(R) (2000).

\end{thebibliography}
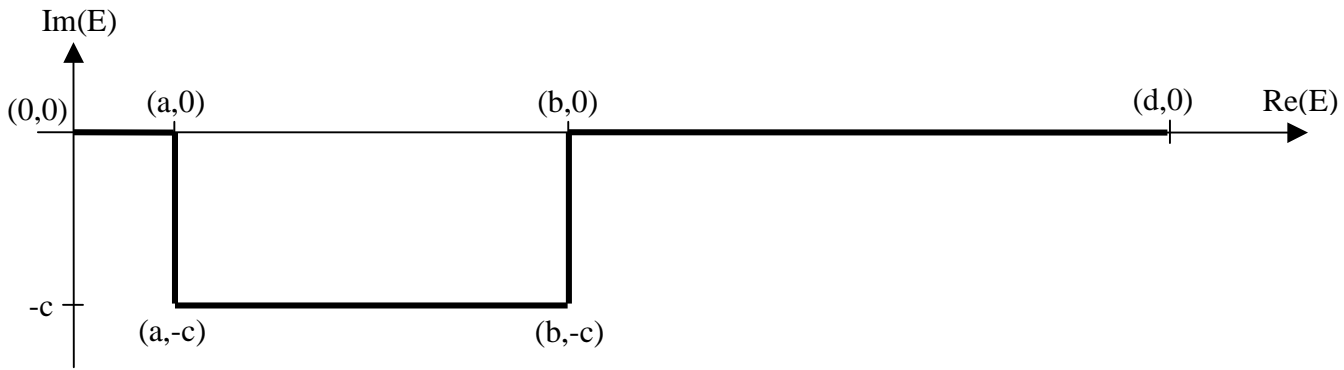
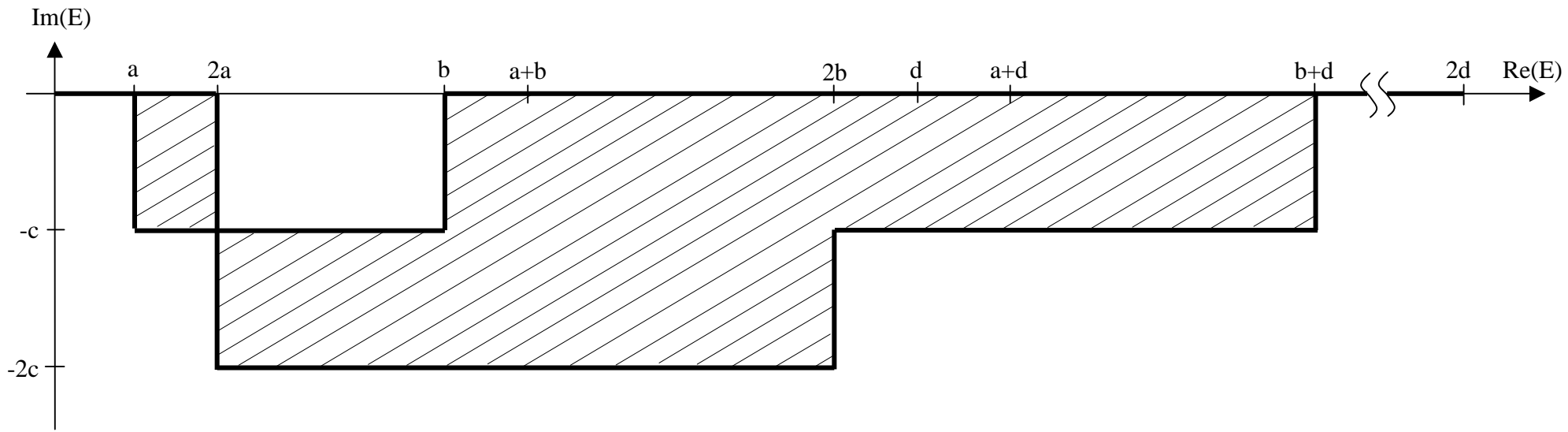
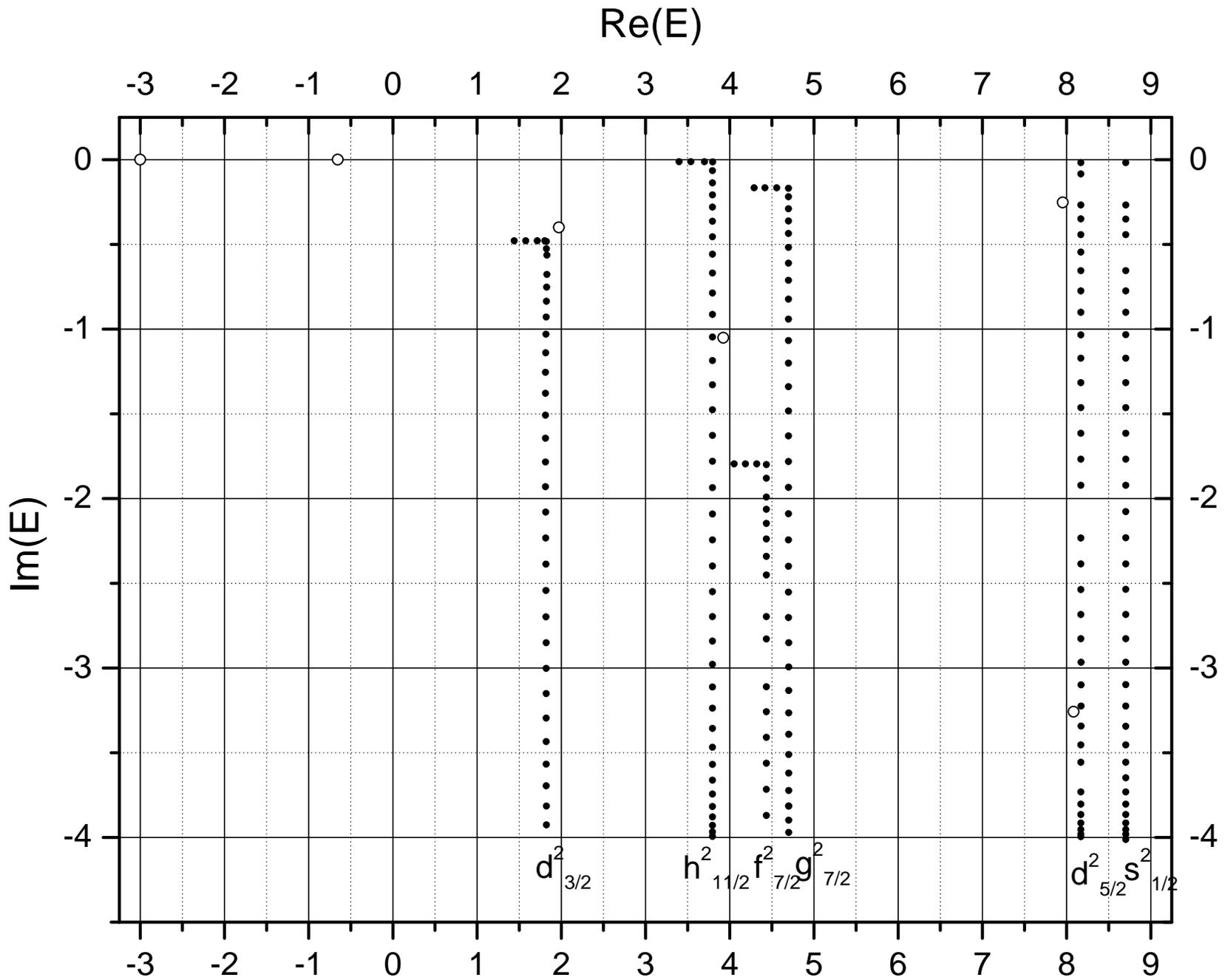
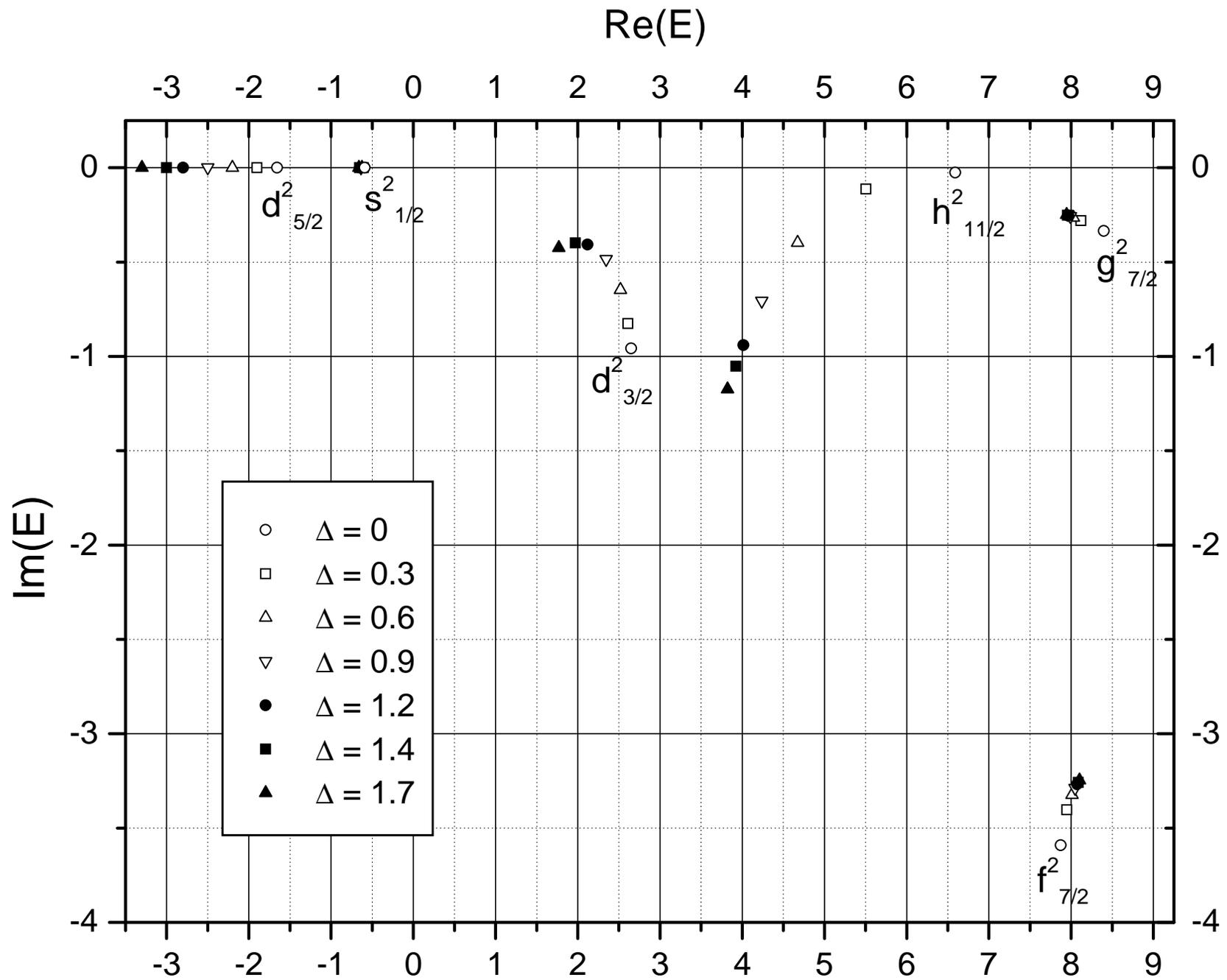
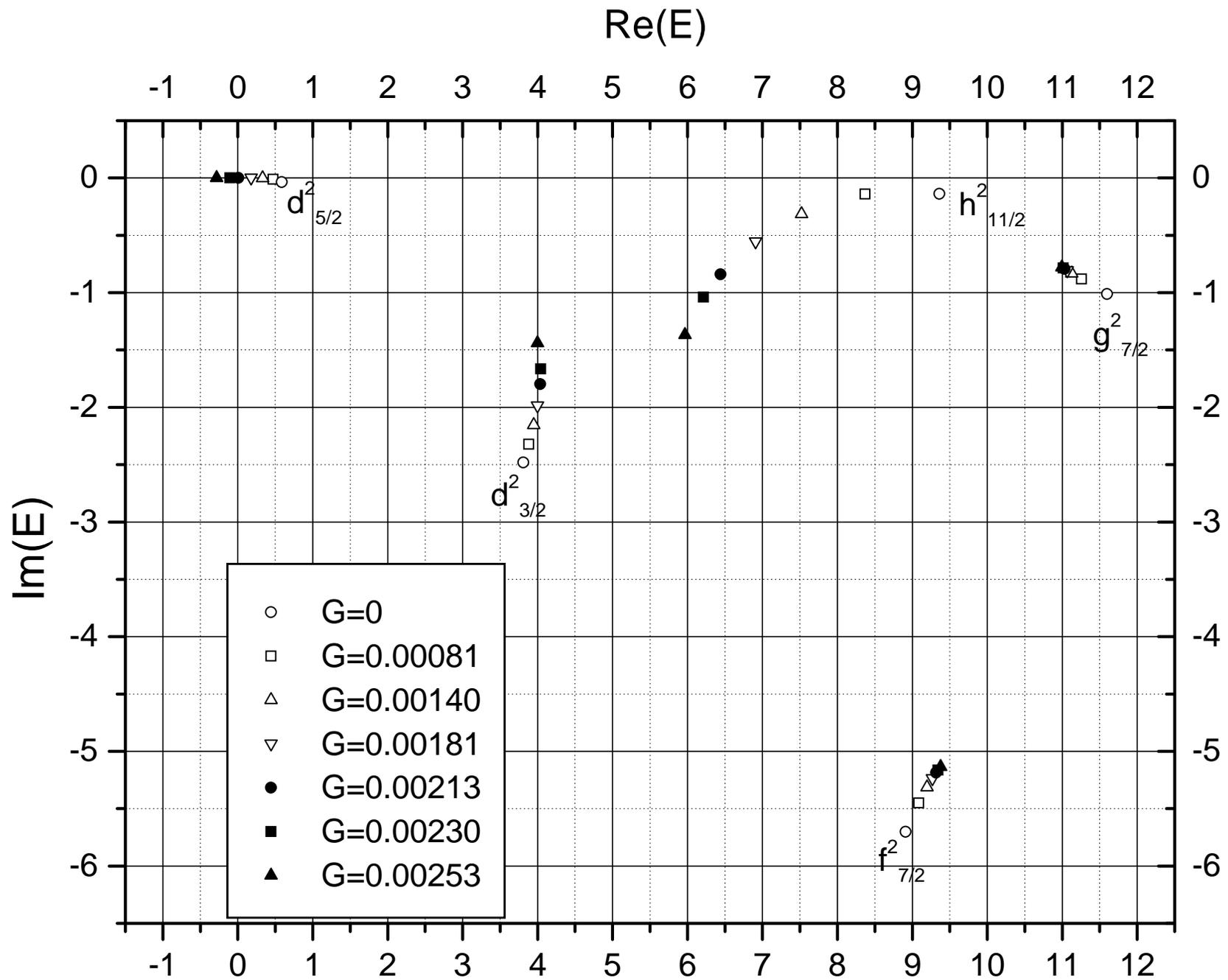
\end{document}